\documentclass[intlimits,twoside,a4paper]{article}
\usepackage[cp1251]{inputenc}
\usepackage[eqsecnum]{cmpj3}
\usepackage{bm}
\usepackage{soul,xcolor}

\issue{2023}{26}{1}{13505}
\doinumber{10.5488/CMP.26.13505}

\title[Interplay of Kekul\'e distortions]%
{Interplay of Kekul\'e distortions and laser fields in graphene%
\thanks{We dedicate this work to Bertrand in the celebration of his 60th anniversary.}}
\author[A. L\'{o}pez, F. Mireles]
{{A. Lopez\orcid{0000-0002-1220-6036}\refaddr{label1}\thanks{Corresponding author: \email{alexlop@espol.edu.ec}.}},
        {F. Mireles\orcid{0000-0003-0506-0793}\refaddr{label2}\thanks{\email{fmireles@cnyn.unam.mx}}}}
\addresses{
\addr{label1}Escuela Superior Polit\'ecnica del Litoral, ESPOL, Departamento de F\'isica, Facultad de Ciencias Naturales y Matem\'aticas, Campus Gustavo Galindo
 Km. 30.5 V\'ia Perimetral, P. O. Box 09-01-5863, Guayaquil, Ecuador
\addr{label2} Departamento de F\'isica, Centro de Nanociencias y Nanotecnolog\'ia, Universidad Nacional Aut\'onoma de M\'exico,
Apartado Postal 14, 22800 Ensenada, Baja California, M\'exico}

\Keywords{graphene, Landau levels, Kekul\'e distortion, Floquet theory, polarization} 

\date{Received November 8, 2022, in final form November 28, 2022}

\begin{document}
\maketitle

\begin{abstract}
We study the interplay between a Kekul\'e patterned distortion in monolayer graphene and a monochromatic laser irradiation, focusing in the long wavelength approximation of its Landau level structure. Exploiting the symmetries of the system, we calculate in the static regime an exact analytical solution for the energy spectrum and its eigenstates, which in turn allows us to find close expressions for the polarizations in the valley and pseudospin degrees of freedom. We find that due to the valley-momentum coupling, the valley polarization exhibits a distinct magnetic response for the two $K$-valley components. Owing to that, the  introduction of the radiation field mixes the Landau levels, and it is shown that by tuning the system to resonance leads to a larger valley-polarization response as compared to the non-resonant scenario.

\printkeywords
\end{abstract}

\section{Introduction}\label{section1}
Since the experimental realization of graphene~\cite{geim0,geim1,geim2} and the related honeycomb lattice two-dimensional materials, such as silicene~\cite{silicene1,silicene2,silicene3}, borophene, dichalcogenides, among others,  a plethora of proposals to engineer and harness their electronic and optical properties have emerged. 
These two-dimensional materials have also emerged as platforms for the realization of topological phases of matter within the static scenario \cite{VonKlitzing1980,Haldane1988,Bernevig2006,Koenig2007,Hasan2010,Kane2005}. The topological insulating phases occur whenever the system exhibits gapless states that conduct along the boundaries of the two dimensional material. These nontrivial topological properties have also been realized in one~\cite{SSH,Kitawaga2013} and three dimensional systems~\cite{3DTI}. Within the driven fields regime~\cite{Grifoni1998,Chu2004,Dora2009, Oka2009}, the emergence of Floquet topological insulators has been put forward. In this case, the systems in a topologically trivial state in the static regime are driven to nontrivial topological phases by means of periodically driven interactions \cite{Lindner2011,Auerbach2011, Kitagawa2010,Kitagawa2011, Foa2011,Platero2013,Lopez2015}. Thus, in a Floquet topological insulator, the generation of the boundary gapless edge states emerges as a consequence of the topological modifications introduced in the energy band spectrum. Interestingly, these topological phases  have also  been  experimentally realized in optical lattices~\cite{ReviewOpticalLattices}. Moreover, the photonic experimental realization of Floquet topological insulators is experimentally realized in reference \cite{Rechtsman2013} by means of a photonic lattice, exhibiting topologically protected transport of visible light on the lattice edges. Additional theoretical work on the optical nonlinear effects in solids~\cite{Marimoto2016} shows that these nonlinear effects can also be properly described within the framework of Berry curvatures.

Apart from periodically driven interactions, other means to engineer the topological properties of low dimensional materials have been proposed in the literature. One interesting realization consists in the generation of a Kekul\'e distortion~\cite{Chamon2000,Chamon2007,Chamon2008} in which the bonding parameters can be modified. This in turn implies that the size of the unit cell is tripled due to the emergent dimerization which amounts to the coupling points in the Brillouin zone which are separated by $\vec{G}=\vec{K}_+-\vec{K}_-$, with $\vec{K}_\pm$ being the Dirac points in graphene. This distortion in turn, results in the merging of the two Dirac cones at the center of the Brillouin zone, producing either a gap (Kek-O) or a superposition of two cones with different Fermi velocities (Kek-Y)~\cite{Montambaux,Gutierrez2016,Liu2017,Beneventano,gamayun}, depending upon the bond length between C-C atoms. The physics of the merging of the Dirac cones has also attracted  attention, as it has been shown that it could lead to semi-Dirac materials in which, the charge carriers possess an anisotropic energy spectrum as their long wavelength effective Hamiltonian consists of a combination of linear and quadratic dispersion relations along two perpendicular directions in the plane~\cite{Wang2018,Gonzalez-Arraga2018,higuchi,MontambauxPRL2013}. In addition, the Kekul\'e distortions can also lead to valley engineering by strain whose strength could serve as control parameter for intervalley scattering processes~\cite{Andrade2019}. Upon extending the study to multilayer systems, there were proposals for bilayer Kekul\'e distorted graphene in which the quasiparticle's masses can be tuned electrostatically. This in turn might lead to versatile platforms for valleytronics with ``multi flavor'' models~\cite{Ruiz-Tijerina}. We refer the interested reader to the review~\cite{Vozmediano} which presents a comprehensible treatment of the role of strain induced distortions in graphene.  
 In a recent work, Mojarro et al.~\cite{Ramon2020} analyze the interplay of Kekul\'e  distortions and electromagnetic radiation perpendicular to the monolayer graphene sample. One natural extension of the model presented in reference~\cite{Ramon2020} is to assess what are the consequences of the interplay of radiation fields and Kekul\'e distortions in the Landau level structure of monolayer graphene. Previous works  considered the interplay in monolayer graphene of quantizing magnetic fields and monochromatic radiation, analyzing the emergent Floquet-Hofstadter spectrum ~\cite{Du2018,Wackerl2019} whereas further works explored the light-induced anomalous Hall effect~\cite{Sato2019}, providing a microscopic theory to explain the underlying mechanisms arising from the light-matter interactions. This light induced anomalous Hall effect in graphene was recently  experimentally realized~\cite{McIver2020}, showing that the dependence of the effect on
a gate potential used to tune the Fermi level reveals multiple features that reflect a Floquet-engineered topological band structure. Further theoretical work~\cite{Foa2021} shows that the chirality of light permits switching on and off  the Hall edge conductance, providing additional control of topologically protected transport.    

To explore the issue of the role of radiation fields and Kekul\'e distortions and to provide a physical insight into this scenario is the aim of this work. Therefore, we theoretically analyze the dynamical manipulation of the Landau level structure of charge carriers on Kekul\'e distorted graphene when a periodically driven radiation field is applied perpendicular to the sample. As  Mojarro et al. show, only the so-called KeK-Y bond pattern couples to the radiation field in the valley degree of freedom. Hence, we focus on this configuration. To deal with the driven regime, we make use of Floquet's theorem to recast the dynamics in an explicitly time-independent fashion, providing an analytical description of the driven evolution of the relevant physical quantities. 
The paper is organized as follows. In section~\ref{sec1} we present the model for the long wavelength approximation of Landau levels induced  on a Kekul\'e distorted honeycomb lattice in presence of circularly polarized electromagnetic radiation, perpendicularly incident upon the sample. Taking advantage of the rotational invariance of the system, we derive the exact effective time-independent Floquet Hamiltonian. Next, in section \ref{sec2} we present and discuss the main results of the manuscript whereas in section \ref{sec4} we present the concluding remarks. Additional calculations are given in the appendix. 

 \section{Model}\label{sec1}
We study the long wavelength effective Hamiltonian for spinless Dirac fermions on a distorted honeycomb lattice with a Kekul\'e bond pattern. The physical manifestation of the Kekul\'e distortion implies that the $K$ and $K'$ Dirac points or valleys are coupled by the wave vector $\vec{G} = \vec{K}_+ - \vec{K}_-$ of the Kekul\'e bond texture and they are folded onto the center of the superlattice Brillouin zone. For the static scenario, we consider the Landau levels in monolayer graphene with a Kekul\'e-Y bond pattern when the sample is subjected to a perpendicular static quantizing magnetic field $B$ and to monochromatic radiation circularly polarized frequency $\omega$. Within these considerations at the $K$ Dirac point,  the Hamiltonian reads $\mathcal{H}(t)=\mathcal{H}_0+\mathcal{V}(t)$.  The first term describes a static
 contribution that in the long wavelength approximation is given explicitly by~\cite{gamayun,Ramon2020}
\begin{equation}
\mathcal{H}_0=v_{\tau}\tau_{0}\otimes\vec{\pi}\cdot\vec{\sigma}+v_\tau\vec{\pi}\cdot\vec{\tau}\otimes\sigma_0, 
\end{equation}
where  $\vec{\pi}=\vec{p}+e\vec{A}$, with $-e$ being the electron charge, $(p_x,p_y)$ being the momentum measured from the $K$ Dirac point, $\vec{A}$ is the static vector potential associated to a perpendicular magnetic field $\nabla\times\vec{A}=\vec{B}$, and $\vec{\sigma}$ ($\vec{\tau}$) represents a vector of Pauli matrices in the sublattice (valley) degree of freedom, whereas $\sigma_0$ and $\tau_0$ represent the $2\times2$ identity matrix, in the corresponding degree of freedom.  Finally, $v_\sigma=v_F$ for pristine graphene, whereas $v_\tau=v_F\Delta_0$ is the corrected Fermi velocity arising from the valley coupling due to the Kekul\'e distortion, within the approximation regime $\Delta_0\ll1$. Before dealing with the radiation effects, we remark that we can find an exact analytical solution to the spectrum and eigenstates of $\mathcal{H}_0$ by means of a different approach as the one presented in \cite{gamayun}. As it is discussed in reference \cite{gamayun}, the static magnetic field implies that the valley and pseudospin contributions do not commute with each other, and yet the spectrum can be solved exactly. To solve the Schr\"odinger equation 
$\mathcal{H}_0|\Psi\rangle=E|\Psi\rangle$, we define the raising and the lowering operators
\begin{eqnarray}
&a=\frac{\ell_B(\pi_x-\ri\pi_y)}{\sqrt{2}\hbar},&\nonumber\\ 
&a^\dagger=\frac{\ell_B(\pi_x+\ri\pi_y)}{\sqrt{2}\hbar},& 
\end{eqnarray}
and write down the static Hamiltonian as
\begin{equation}\label{h0}
 \mathcal{H}_0=\hbar\omega_c\left(
\begin{array}{cccc}
 0&a&\Delta_0 a&0\\
 a^\dagger& 0&0&\Delta_0 a\\
\Delta_0a^\dagger&0&0& a\\
0&\Delta_0a^\dagger& a^\dagger&0
\end{array}
\right),
\end{equation}
with $\omega_c=\sqrt{2}v_F/\ell_B$ describing the cyclotron frequency. In addition, $v_F\approx10^6$ m/s is the Fermi velocity, whereas $\ell_B^{-2}=eB/\hbar$ 
is the magnetic length defined in terms of the strength of the quantizing magnetic field $B$ and $-e$ is the electron charge.  
 In reference~\cite{gamayun}, the spectrum of the Hamiltonian in equation (\ref{h0}) is solved  using  an artificial site energy term of small value that  afterwards  is made to vanish, so that the solutions are shown to be independent of the on site energy parameters. Alternatively, one can instead exploit  the symmetries of the system to solve the Landau level spectrum in a more physically appealing form. To this end, we notice first that the static Hamiltonian commutes with the number operator
\begin{equation}\label{nz}
\mathcal{N}_a=a^\dagger a\tau_0\otimes\sigma_0+\frac{\sigma_z+\tau_z}{2},
\end{equation} 
whereby we can look for simultaneous eigenstates of this number operator and the static Hamiltonian $\mathcal{H}_0$. If we write
\begin{equation}\label{ansatz}
|\Psi_{\sigma\tau}(n)\rangle=\left(
\begin{array}{c}
 A^\tau_n|n-1\rangle\\
 \sigma B^\tau_n|n\rangle\\
 C^\tau_n|n\rangle\\
 \sigma D^\tau_n|n+1\rangle
\end{array}
\right),
\end{equation} 
with the states $|n\rangle$ being eigenstates of the Hermitian operator $n=a^\dagger a$, i.e.,  $n=a^\dagger a|n\rangle=n|n\rangle$, it is easily verified that $\mathcal{N}_a|\Psi_{\sigma\tau}(n)\rangle=n|\Psi_{\sigma\tau}(n)\rangle$. In absence of quantizing magnetic field, the number operator can be identified with the total angular momentum operator whose conservation is a consequence of the rotational invariance of the graphene Hamiltonian. 
 In the limit of a vanishing Kekul\'e parameter $\Delta_0\rightarrow0$, the expansion coefficients must satisfy $B^-_n= A^-_n$ and $C^+_n= D^+_n$, along with $B^+_n= A^+_n=C^-_n= D^-_n=0$, as we should recover the valley degenerate solutions for the Landau levels.  
\begin{figure}[!t]
	\begin{center}
\includegraphics[height=5.3cm]{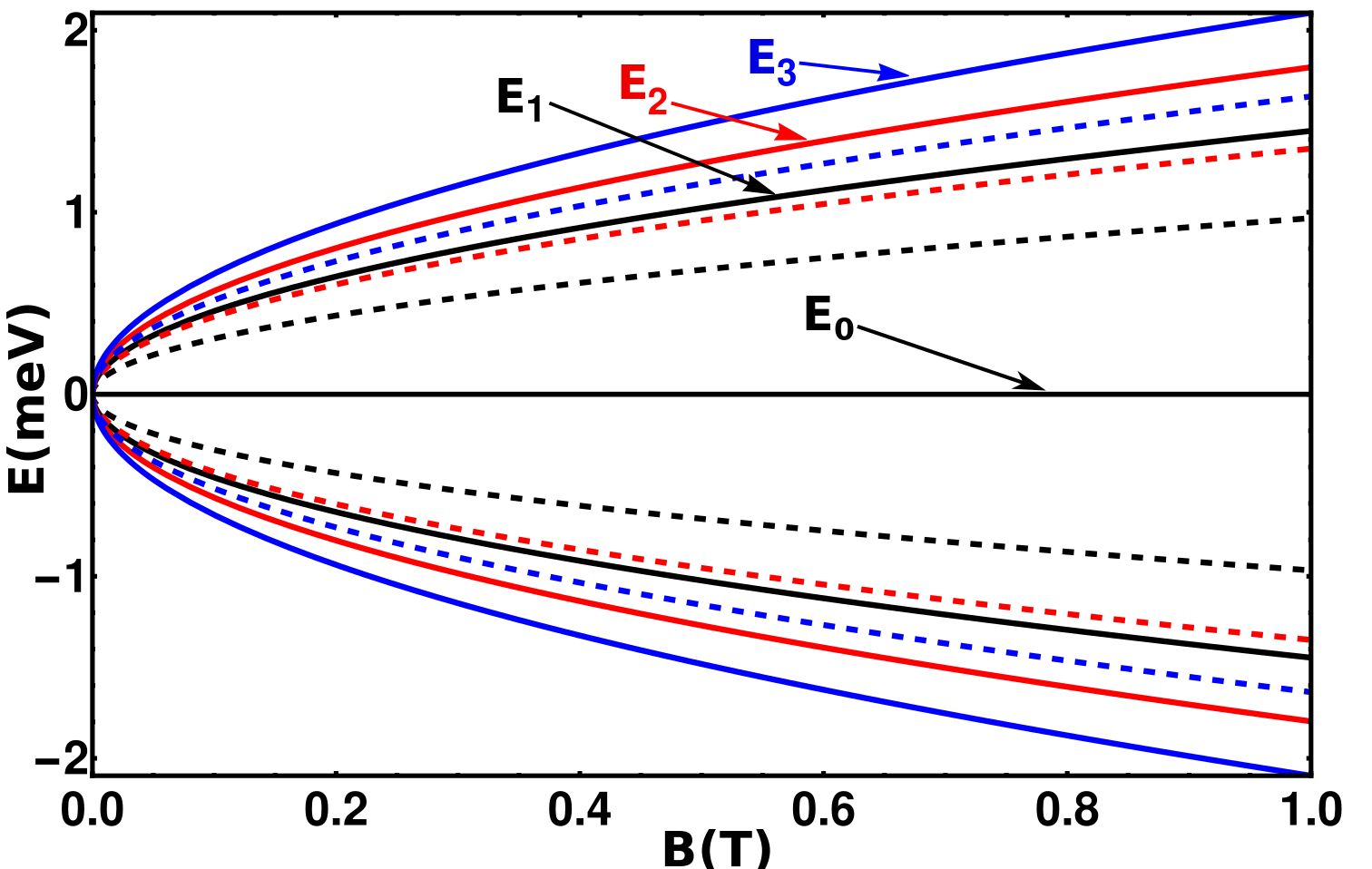}
\caption{(Colour online) Magnetic field dependence of the first three Landau levels with the continuous (dashed) lines corresponding to the valley index $\tau=1$ ($\tau=-1$) as given by equation (\ref{static}). We notice the valley splitting even for a small Kekul\'e parameter $\Delta_0=0.1$ (see the main text). }  
\label{fig1}  
\end{center}
\end{figure}
 Using this state, and after some algebraic manipulations, the energy spectrum reads  $E_{\sigma\tau}(n)=\sigma E^\tau_{n}$ (see appendix for details),  where we have defined:
\begin{equation}\label{static}
 E^\tau_{n}=\hbar\omega_c\sqrt{\left(n+\frac{1}{2}\right)(1+\Delta_0^2)+\frac{1}{2}\tau\sqrt{16n(n+1)\Delta_0^2+(1+\Delta_0^2)^2}}\quad n = 1, 2, 3,\ldots\,.
\end{equation}
The indexes take on the values $\sigma,\tau=\pm1$, whereas $n=0,1,2...$ labels the different Landau levels, but one should note that for  $n=0$, equation~(\ref{static}) does not apply (this case is treated separately below). This energy spectrum (figure~\ref{fig1}) is equivalent to that obtained in reference \cite{gamayun} but we avoid the need to introduce a spurious onsite energy term which they use to obtain their result.
From equation (\ref{static}), we observe that in the vanishing distortion parameter limit $\Delta_0\rightarrow 0$, we recover the degenerate spectrum
\begin{equation}
(E^{-}_n)^2=(E^{+}_{n-1})^2=n\hbar^2\omega_c^2. 
\end{equation}
After some lengthy algebraic operations, we get for the coefficients of the eigenstate in equation (\ref{ansatz}) the expressions for $\tau=-1$ and $n\ne0$:
\begin{eqnarray}\label{coefficients1}
B^-_n&=&
\frac{\epsilon_n^{-}}{\sqrt{n}}\left[\frac{(\epsilon^{-}_n)^2-(n+1)-n\Delta_0^2}{(\epsilon^{-}_n)^2-(n+1)(1-\Delta_0^2)}\right]A_n^-
\equiv b^-_n A^-_n,\nonumber\\
C^-_n&=&\Delta_0\left(\frac{\sqrt{n}\epsilon_n^-+(n+1)b_n^-}{(\epsilon_n^-)^2-(n+1)}\right)A_n^-\equiv c^-_n A^-_n,\nonumber\\
D^-_n&=&\Delta_0\sqrt{n+1}\left(\frac{\epsilon_n^-b_n^-+\sqrt{n}}{(\epsilon_n^-)^2-(n+1)}\right)A_n^-\equiv d_n^- A_n^-,
\end{eqnarray}
whereas those corresponding to $\tau=+1$ and $n\ne0$, read as
\begin{eqnarray}\label{coefficients2}
A^+_n&=&\Delta_0\sqrt{n}\left(\frac{\epsilon_n^+c_n^++\sqrt{n+1}}{(\epsilon_n^+)^2-n}\right)D_n^+\equiv a^+_n D^+_n,\nonumber\\
B^+_n&=&\Delta_0\left(\frac{\sqrt{n+1}\epsilon_n^++nc_n^+}{(\epsilon_n^+)^2-n}\right)\equiv b^+_n D^+_n,\nonumber\\
C^+_n&=&\frac{\epsilon_n^{+}}{\sqrt{n+1}}\left[\frac{(\epsilon^{+}_n)^2-n-(n+1)\Delta_0^2}{(\epsilon^{+}_n)^2-n(1-\Delta_0^2)}\right]D_n^+
\equiv c^+_n D^+_n,
\end{eqnarray}
and the remaining coefficients $A^-_n$ and $D^+_n$  being global factors, would not be needed for the calculation of physical properties. In the explicit calculations, we use $a^-_n=d^+_n=1$. Here, we have also introduced the dimensionless quantity $\epsilon^\pm_{n}=E^\pm_{n}/\hbar\omega_c$. The special case of $n=0$, corresponding to a valley degenerate zero-energy solution  $E_{\sigma\tau}(0)=0$, with (normalized) eigenstate, reads as:
\begin{equation}\label{zero}
|\Psi_0\rangle=-\frac{1}{\sqrt{1+\Delta_0^2}}\left(
\begin{array}{c}
 0\\
 -|0\rangle\\
 \Delta_0|0\rangle\\
 0
\end{array}
\right).
\end{equation} 
\section{Results}\label{sec2}
We now explore the physical consequences of the interplay of Kekul\'e and electromagnetic monochromatic radiation circularly polarized. First, we analyse the static polarization effects and afterwards we extend the analysis to the driven scenario.
\subsection{Static regime}
\begin{figure}[!t]
\centering
\includegraphics[height=4cm]{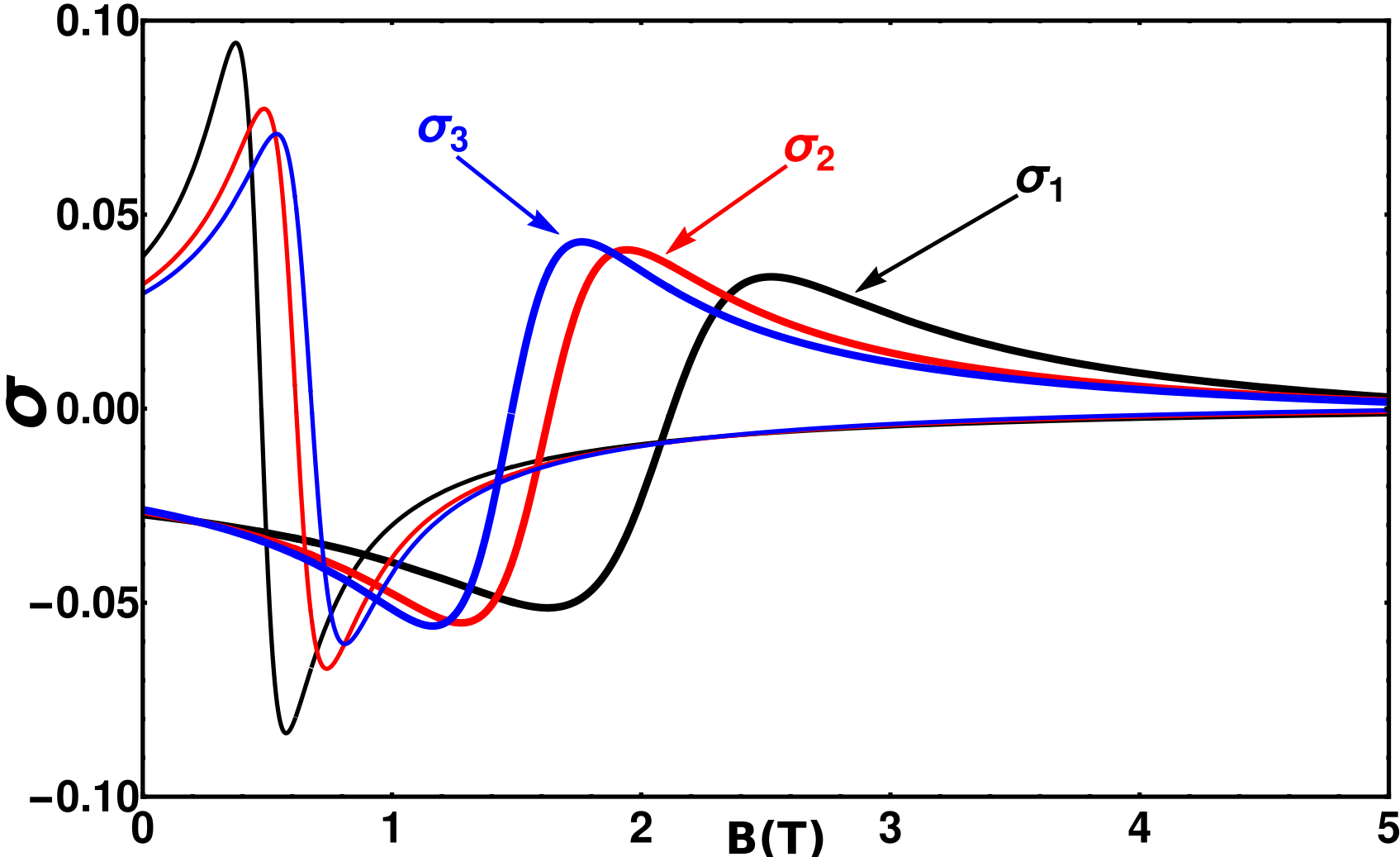}
\includegraphics[height=4cm]{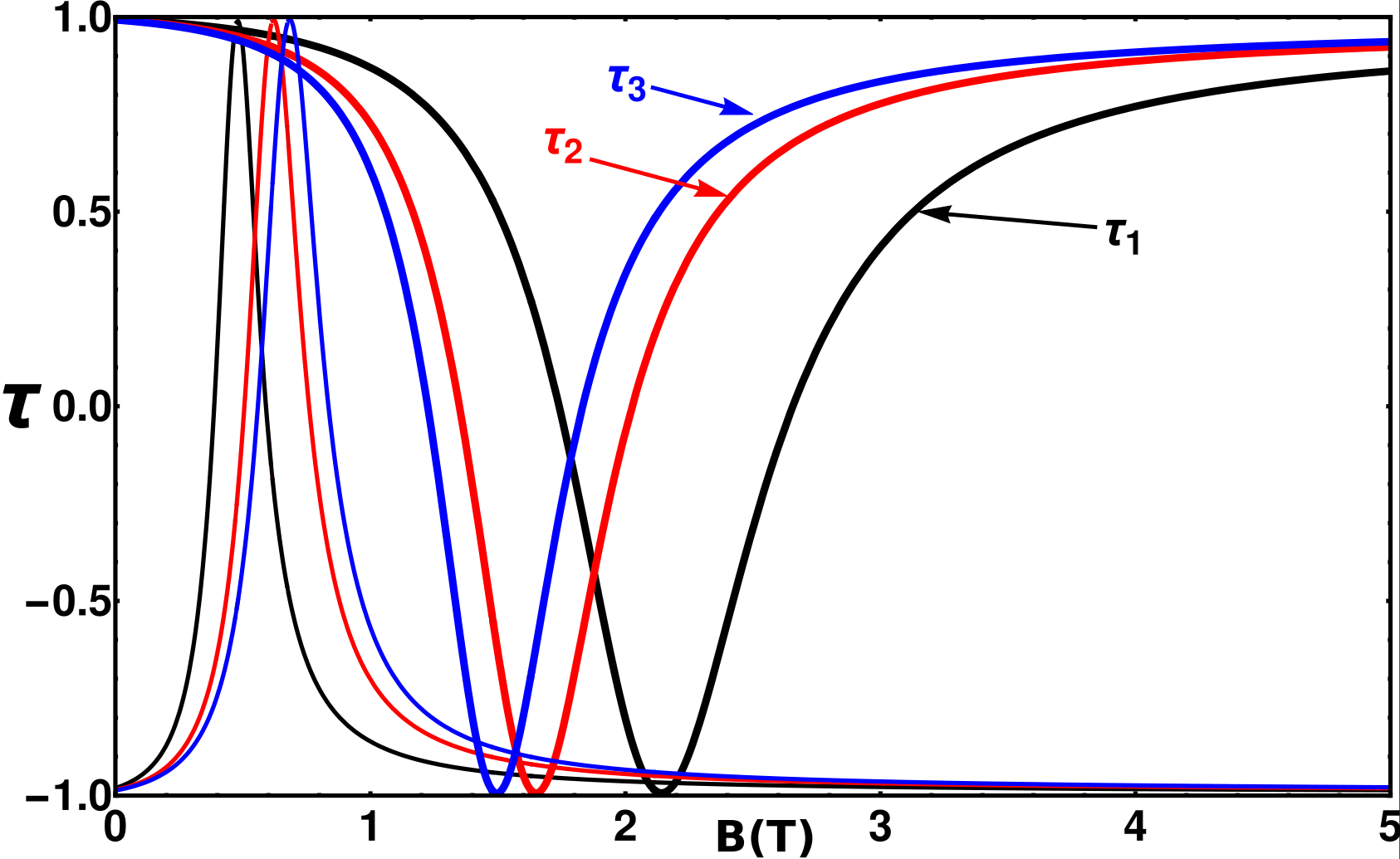}
\caption{(Colour online) Left-hand panel: Out of plane static pseudospin polarization for the first three Landau levels with the thick (thin) lines corresponding to valley index $\tau=1$ ($\tau=-1$). Right-hand panel: Valley polarization as given by equations (\ref{sigma-tau}). We notice the valley sppliting even for a small Kekul\'e parameter $\Delta_0=0.1$ (see the main text). }  
\label{fig2}    
\end{figure}

By using the coefficients in equation (\ref{coefficients1}) and (\ref{coefficients2}), we can evaluate the pseudospin polarization $\langle\sigma_z\rangle^\pm_n=\langle\Phi^\pm_n|\tau_0\otimes\sigma_z|\Phi_{n}^{\pm}\rangle$ and the valley polarizations $\langle\tau_z\rangle^{\pm_n}=\langle\Phi{^\pm_n}|\tau_z\otimes\sigma_0|\Phi{^\pm_n}\rangle$, for any state $n\neq0$ with $\pm$ denoting here the Landau level associated to the electron/hole bands. The pseudospin and valley polarization per Landau level are  given by 
\begin{eqnarray}\label{sigma-tau}
\langle\sigma_z\rangle_n^-&=&\frac{1-|b^-_n|^2+|c^-_n|^2-|d^-_n|^2}{1+|b^-_n|^2+|c^-_n|^2+|d^-_n|^2},\nonumber\\
\langle\sigma_z\rangle_n^+&=&\frac{|a^+_n|^2-|b^+_n|^2+|c^+_n|^2-1}{1+|a^+_n|^2+|b^+_n|^2+|c^+_n|^2},\nonumber\\
\langle\tau_z\rangle_n^-&=&\frac{1+|b^-_n|^2-|c^-_n|^2-|d^-_n|^2}{1+|b^-_n|^2+|c^-_n|^2+|d^-_n|^2}, 
\nonumber\\
\langle\tau_z\rangle_n^+&=&\frac{|a^+_n|^2+|b^+_n|^2-|c^+_n|^2-1}{1+|a^+_n|^2+|b^+_n|^2+|c^+_n|^2},
\end{eqnarray}
which in the limit of the vanishing Kekul\'e distortion lead to $\langle\sigma_z\rangle_n^\pm=0$ and $\langle\tau_z\rangle_n^\pm=\pm1$.
In the left-hand (right-hand) panel of figure~\ref{fig2} we show the out of
plane static pseudospin (valley) static polarization for the first three
Landau leveles $n=1,2,3$, which show the valley splitting, even for a small
Kekul\'e parameter $\Delta_0=0.1$.
 We also find that for $n=0$, these polarizations simply read as:
\begin{eqnarray}
\langle\sigma_z\rangle_0&=&\frac{-1+\Delta_0^2}{1+\Delta_0^2},\nonumber\\
\langle\tau_z\rangle_0&=&\frac{1-\Delta_0^2}{1+\Delta_0^2}, 
\end{eqnarray}
and we verify that $\langle\sigma_z\rangle_0=-\langle\tau_z\rangle_0$ is a direct consequence of the condition of null excitations in the $n=0$ Landau level (LL), i.e., $\mathcal{N}_a|\Psi_0\rangle=0$. 
The explicit polarization dependence on $\Delta_0$  could be interpreted as a supporting argument to the initial statement that the bonding pattern couples these two degrees of freedom.

\subsection{Dynamic regime}
 We now consider the role of the radiation field in modulating the LL structure of the Kekul\'e distorted graphene monolayer. To this end, we start from the ordinary dipolar interaction term $-e\vec p\cdot\vec A(t)$, which can be introduced to the tight-binding Hamiltonian via the Peierls 
substitution. Under the long wavelength regime of interest we can evaluate, near the Dirac points, that the effects of the driving field are incorporated through the minimum
coupling prescription, leading to the time-dependent term  
\begin{equation}
\mathcal{V}(t)=ev_\sigma\vec{\sigma}\cdot\vec{A}(t)+ev_\tau\vec{\tau}\cdot\vec{A}(t),
\end{equation} 
thus, the total Hamiltonian reads
\begin{equation}\label{time}
\mathcal{H}(t)=\mathcal{H}_0+\mathcal{V}(t).
\end{equation}
For a monochromatic circularly polarized radiation field  impinging perpendicular to the graphene monolayer, the vector potential can be chosen as $\vec{A}(t)=A(\cos\omega t, \sin\omega t)$, with $A=\mathcal{E}/\omega$. 
This makes the total Hamiltonian  periodic in time, $\mathcal{H}(t+T)=\mathcal{H}(t)$, with $T=2\piup/\omega$ the period of oscillation of the driving field. 
The parameters $\mathcal{E}$ and $\omega=2\piup/T$ correspond to the amplitude and frequency of the radiation field, respectively. The electric field is given in turn by the standard relation $\vec{\mathcal{E}}(t)=-\partial_t \vec{A}(t)$. 
Then, the time-dependent contribution reads
\begin{equation}\label{bilayerint}
\mathcal{V}(t)=\xi\left(
\begin{array}{cccc}
 0&\re^{-\ri\omega t}&\Delta_0 \re^{-\ri\omega t}&0\\
 \re^{\ri\omega t}& 0&0&\Delta_0 \re^{-\ri\omega t}\\
\Delta_0 \re^{\ri\omega t}&0&0& \re^{-\ri\omega t}\\
0&\Delta_0 \re^{\ri\omega t}& \re^{\ri\omega t}&0
\end{array}
\right),
\end{equation} 
where we have introduced the effective light-matter coupling strength $\xi=e\mathcal{E}v_F/\omega$. 
 We assume that the beam radiation spot is large enough compared to the lattice spacing, and we can neglect any spatial variation.

We can perform a time-dependent unitary transformation $\mathcal{P}(t)=\re^{-\ri\mathcal{N}_a\omega t}$, given explicitly as $\re^{-\ri n\omega t}\text{diag}[\re^{-\ri\omega t},1,1,\re^{\ri\omega t}]$, in the form
\begin{equation}
\mathcal{H}_F=\mathcal{P}^\dagger(t)\mathcal{H}(t)\mathcal{P}(t)-\ri\hbar\mathcal{P}(t)\partial_t\mathcal{P}(t),
\end{equation}
after which we get a time-independent Floquet Hamiltonian~\cite{Grifoni1998,Chu2004}
\begin{equation}\label{heff}
 \mathcal{H}_{F}=\hbar\omega_c\tau_0\otimes(a^\dagger\sigma_-+a\sigma_+	)+\hbar\omega_c\Delta_0(a^\dagger\tau_-+a\tau_+	)\otimes\sigma_0-\mathcal{N}_a\hbar\omega+\xi\tau_0\otimes\sigma_x+\xi\Delta_0\tau_x\otimes\sigma_0.
 \end{equation}
 Up till now, the results are analytically exact and we could solve numerically the resulting time-independent Schr\"odinger equation $\mathcal{H}_{F}|\Phi_F(t)\rangle=\varepsilon|\Phi_F(t)\rangle$ to obtain the quasienergy spectrum $\varepsilon$ and Floquet eigenstates $|\Phi_F(t)\rangle$.
 Some comments are in order here. Within the approach of  Fourier mode expansion to numerically solve the Floquet Hamiltonian, the sidebands associated to  the periodicity of the quasienergy spectrum yield $\varepsilon_N\,\text{mod}\,{\hbar\omega}$. On the other hand, from the obtained effective Floquet Hamiltonian given in (\ref{heff}) we notice that the role of the radiation field is actually to couple the different LLs. Indeed, for a given value of the LL index, the factor $\mathcal{N}_a\hbar\omega$ would represent the discrete energy translations of  quasienergies. Moreover, we clearly see that the last two terms of (\ref{heff})  do not commute with the static Hamiltonian; hence, they would produce the level mixing which no longer renders the index $n$  a good quantum number. This can be explicitly shown by expressing the Floquet Hamiltonian in the static eigenbasis (\ref{ansatz}):
 \begin{equation}
 \langle\Phi_{n'}^{\sigma '\tau'}|\mathcal{H}_F|\Phi_{n}^{\sigma \tau}\rangle=(E^{\sigma\tau}_n-n\hbar\omega)\delta_{nn'}\delta_{\sigma\sigma'} \delta_{\tau\tau'}+V_{nn'},\nonumber
 \end{equation} 
with the matrix elements of the coupling term 
\begin{eqnarray}
V_{nn'}&=&\xi\sum_{s=\pm1}\frac{(b_n+c_{n+s}d_n)\delta_{n',n+s}}{\sqrt{(1+|b_n|^2+|c_n|^2+|d_n|^2)(1+|b_{n+s}|^2+|c_{n+1}|^2+|d_{n+s}|^2)}}\nonumber\\ 
&+&\xi\sum_{s=\pm1}\Delta_0\xi\frac{(c_n+b_{n+s}d_n)\delta_{n',n+s}}{\sqrt{(1+|b_n|^2+|c_n|^2+|d_n|^2)(1+|b_{n+s}|^2+|c_{n+s}|^2+|d_{n+s}|^2)}}.
\end{eqnarray}
 
However, this approach does not provide a clear picture of the most relevant features of the underlying photoinduced processes. Therefore, in the following section we  introduce physically appealing solutions within appropriate parameter regimes of experimental relevance.
\subsection{Results for the driven regime}
In the previous section we  derived an analytical expression for the Floquet Hamiltonian. To get a further physical insight, now we assess the role of the radiation field in different parameter regimes. We explore the valley and pseudospin polarization dynamics within these different parameter configurations. In the following discussion we use as the energy scale the cyclotron frequency $\omega_c$.
To proceed further, we introduce the shifted ladder operators $b=a+\xi/\hbar\omega_c$, which allow us to rewrite the exact Floquet Hamiltonian as
\begin{equation}\label{shifted}
 \mathcal{H}_{F}=\hbar\omega_c\tau_0\otimes(b^\dagger\sigma_-+b\sigma_+	)+\hbar\omega_c\Delta_0(b^\dagger\tau_-+b\tau_+)\otimes\sigma_0-\mathcal{N}_b\hbar\omega+\lambda\hbar\omega(b^\dagger+b)-\lambda^2\hbar\omega,
 \end{equation}
with $\lambda=\xi/\hbar\omega_c$. In this form, we can assess the role of the different contributions to the Landau level mixing due to the radiation field. 

We now focus on the light-matter coupling analysis. Within the general scenario of laser irradiation, at low frequencies compared to the static bandwidth, the dynamical problem is rather complicated since different Floquet replica (sidebands) might overlap. Indeed, in the general Floquet formulation of the problem, it has been shown that the number of sidebands require to obtain a physically relevant numerical solution of the driven system is to use a number of modes up to $n_{\text{max}}$ which will depend on whether one is in the weak $\xi\ll\hbar\omega$ or strong $\xi\gg\hbar\omega$ coupling limit~\cite{Buttiker1,Buttiker2}. Our solution written in terms of shifted operators given in equation (\ref{shifted}) shows that the leading order light-matter interaction strength that couples the different Landau levels is independent of the frequency. Therefore, the energy corrections (apart from an irrelevant energy shift $-\lambda^2\omega$) would only depend on the amplitude of the radiation field. This in turn might have the advantage of allowing the study of experimental configurations in which the stability of the sample is not frequency dependent in order to avoid damaging the sample~\cite{Cao,Park}.

However, the periodicity of the quasienergy spectrum allows for a resonant behaviour among adjacent Landau levels that turn out to be coupled by the radiation field. Thus, whenever the effective amplitude of the radiation field is smaller than the static bandwidth and the system is out of resonance, we can  approximate the Floquet quasienergy spectrum by means of degenerate perturbation theory. That is, within the weak coupling limit regime of experimental interest, we can use standard time-independent perturbation theory to assess the corrections to the energy spectrum $\Delta E(n)$ for $n\neq0,1$ which up the aforementioned negative global higher-order energy shift $-\lambda^2\omega$, are given by
\begin{equation}\label{corrections}
\Delta E^{\sigma\tau}_n=[\sigma(E^\tau_{n}+E^{-\tau}_{n-1})-\hbar\omega](f^{\sigma\tau}_{n-1})^2+[\sigma(E^\tau_{n}+E^{-\tau}_{n+1})-\hbar\omega](f^{\sigma\tau}_{n+1})^2,
\end{equation}
for which the expressions for the coefficients $f^{\sigma\tau}_{n\pm1}$ are somewhat lengthy, so we provide them in the appendix, along with its derivation.
\begin{figure}
\begin{center}
\includegraphics[height=5cm]{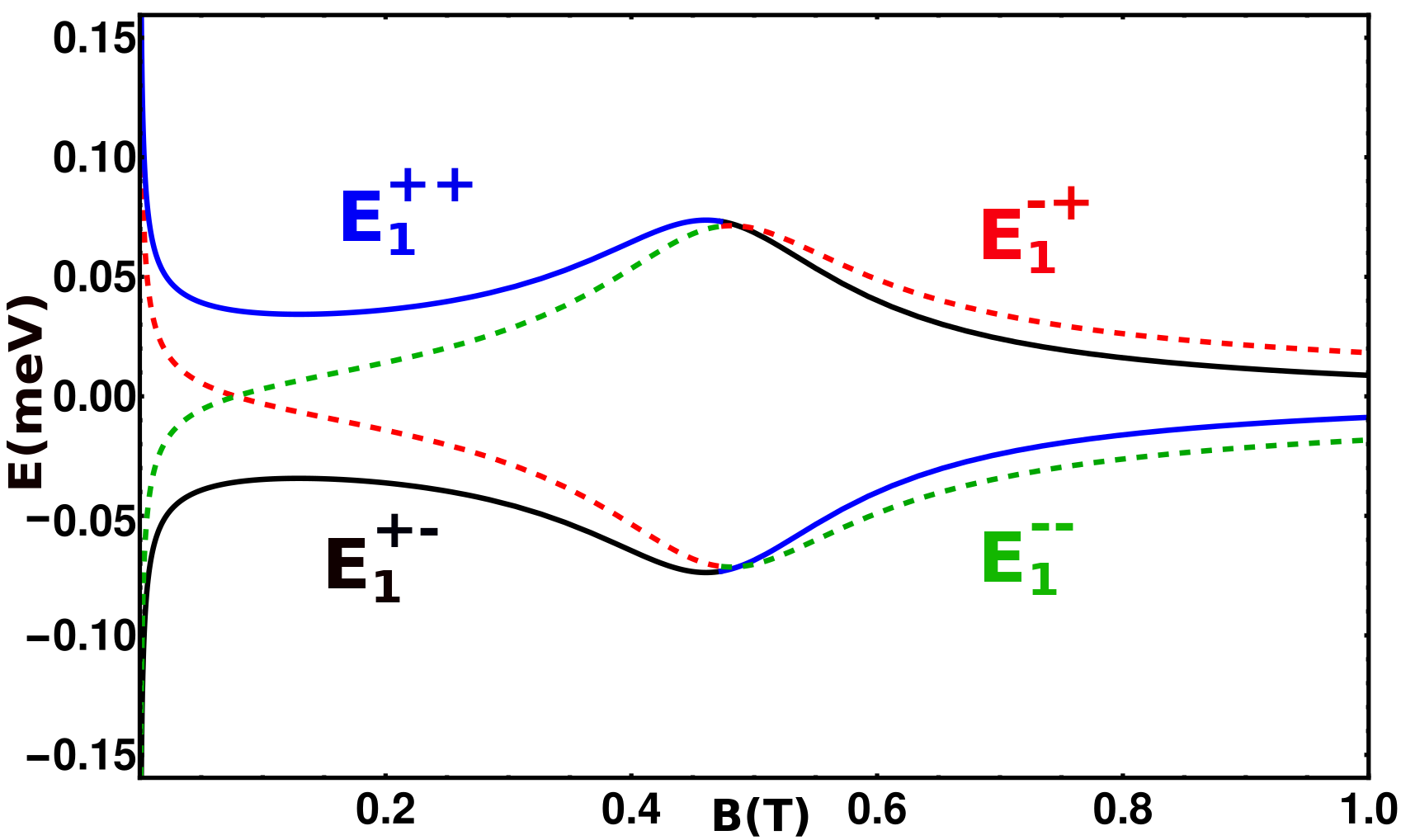}
 \caption{(Colour online) Magnetic field dependence of the resonant energy structure due to the photoinduced mixing of the lowest Landau levels $n=0$ and $n=1$. We have set $e\mathcal{E}\ell_{B_{\text{max}}}=0.1\hbar\omega$, with $\ell_{B_{\text{max}}}$ being the magnetic length at the largest value of the static magnetic field. An asymmetric response of the valley components $\tau=\pm1$ which could be exploited for photoinduced valley modulations (see the main text). }
 \label{fig3}
\end{center}
\end{figure}
Interestingly, only transitions with $\Delta n=\pm1$ are allowed, such that higher order transitions are forbidden. Up to the leading order, the corrected Floquet eigestates read for $n\neq0,1$
\begin{equation}
|\Phi^{\sigma\tau}_n\rangle=|\Psi^{\sigma\tau}_n\rangle +f^{\sigma\tau}_{n-1}|\Psi^{-\sigma,-\tau}_{n-1}\rangle+f^{\sigma\tau}_{n+1}|\Psi^{-\sigma,-\tau}_{n+1}\rangle.
\end{equation}
The energy corrections given in equation (\ref{corrections}) show the standard resonant structure that we now deal with. The energy corrections are determined by analyzing the subspace of zero energy degeneracy between the static $n=0$ zero energy level and the $n=1$ Landau level, which occures whenever $\hbar\omega=\pm E_1^\tau$. This in turn implies that radiation fields with right (left) circular polarization state would pick the $+E_1^\tau$ ($-E_1^\tau$), for any fixed values of the quantizing static magnetic field $B$ and Kekul\'e distortion parameter $\Delta_0$.  
Thus, at resonance these two lowest Landau levels mix to give rise to the leading order quasienergy solution 
\begin{equation}
E_1^{\tau s}=s\frac{e\mathcal{E}\ell_B}{\sqrt{2}}\frac{b_1^\tau-\Delta_0 c_1^\tau}{\sqrt{1+\Delta_0^2}\sqrt{|a_1^\tau|^2+|b_1^\tau|^2+|c_1^\tau|^2+|d_1^\tau|^2}}, 
\end{equation}
with $a_1^-=d_1^+=1$. These resonant energies are modulated both by the static magnetic field and by the amplitude of the radiation field. In figure~\ref{fig3} we plot $E_1^{\tau s}$ as functions of the static magnetic field, where a clear different resonant response for each valley component is observed. This in turn can be traced back to the satic energy spectrum shown in figure~\ref{fig1} where the presence of the Kekul\'e distortion makes the $\tau=-1$ states  get closer to the zero energy lowest Landau level. Thus, we observed a level ``repulsion'' in the most distant energy state that couples more strongly within the resonant regime.
In the degenerate subspace spanned by the two lowest Landau levels, we obtain the Floquet states valid at resonance:
\begin{equation}
|\Phi_1^{\tau s}\rangle=\frac{1}{\sqrt{2}}\left(|\Psi_0\rangle+s|\Psi^\tau_1\rangle\right). 
\end{equation}
This degenerate state is an equally weighted superposition of the static orthogonal states corresponding to the two lowest Landau levels, which can be contrasted to its non-degenerate counterpart
\begin{equation}\label{offresonance}
|\Phi_1^{\tau\sigma}\rangle=|\Psi_1^{\tau\sigma}\rangle+f^{\tau\sigma}_1|\Psi_0\rangle, 
\end{equation}
where
\begin{equation}\label{f1coefficient}
f^{\tau\sigma}_1= \frac{e\mathcal{E}\ell_B}{\sqrt{2(1+\Delta_0^2)(|a_1^\tau|^2+|b_1^\tau|^2+|c_1^\tau|^2+|d_1^\tau|^2)}}\left(\frac{b_1^\tau-\sigma \Delta_0 c_1^\tau}{E_1^\tau-\sigma \hbar\omega}\right).
\end{equation}
Thus, as we  would physically expect, within the off resonant scenario, the relative weight of the Floquet eigenstates mixing the static eigenstates would be much smaller as compared to the resonant regime. 

Moreover, the pseudospin and valley polarizations in the resonant regime are explicitly found to read in terms of the static polarizations as
\begin{eqnarray}
\langle\Phi_1^{\tau s}|\tau_0\otimes\sigma_z|\Phi_{1}^{\tau s}\rangle&=&\frac{1}{2}(\langle\sigma_0\rangle+\langle\sigma^\tau_1\rangle),\nonumber\\ 
\langle\Phi_1^{\tau s}|\tau_z\otimes\sigma_0|\Phi_{1}^{\tau s}\rangle&=&\frac{1}{2}(\langle\tau_0\rangle+\langle\tau^\tau_1\rangle),
\end{eqnarray}
whereas the off resonant expressions are
\begin{eqnarray}
\langle\Phi_1^{\tau\sigma}|\tau_0\otimes\sigma_z|\Phi_{1}^{\tau \sigma}\rangle&=&\langle\sigma^\tau_1\rangle+(f^{\tau\sigma}_1)^2\langle\sigma_0\rangle,\nonumber\\ 
\langle\Phi_1^{\tau\sigma}|\tau_z\otimes\sigma_0|\Phi_{1}^{\tau \sigma}\rangle&=&\langle\tau^\tau_1\rangle+(f^{\tau\sigma}_1)^2\langle\tau_0\rangle,  
\end{eqnarray}
with $f^{\tau\sigma}_1$ being defined in equations (\ref{offresonance}) and (\ref{f1coefficient}).

\section{Summary and conclusions}\label{sec4}

We have addressed the light-matter interaction effects of charge carriers in monolayer graphene with a Kekul\'e bond pattern in the so-called KeK-Y configuration where the radiation field couples to the valley degree of freedom. By restricting ourselves to the long wavelength approximation, we first have shown an exact analytical solution to the static Landau level spectrum for monolayer graphene with a Kekul\'e bond pattern. For this purpose, we have used an alternative approach as that given in reference~\cite{gamayun}. Following the proposal by Mojarro et al.~\cite{Ramon2020}, we have explored the role of monochromatic radiation incident perpendicularly to the graphene monolayer. Interestingly, even for a small value of the bonding parameter, within the weak coupling of the light-matter interaction, a distinctive physical regime emerges in the resonant driven scenario. We have shown that these resonances can be addressed by either right-hand or left-hand polarized radiation fields. In addition, the effective light-matter interaction strength depends only on the radiation electric field amplitude rather than on its frequency. The frequency dependence is addressed through the resonant properties of the driven system. The effective time independent Floquet Hamiltonian shows that only adjacent Landau levels get coupled. This is explicitly analysed in the subspace of the lowest Landau level configuration showing the well known level mixing of degenerate states. Furthermore, the static states get dressed by the driving interaction and a quasienergy band splitting of the zero energy solutions can be exploited to enhance the polarization properties of the KeK-Y graphene system.  
\section*{Acknowledgements}
A. L. thanks Bertrand for more than 17 years of scientific collaboration and friendship in which he has always been an inspiration. F. M. acknowledges the support of DGAPA-UNAM through the project PAPIIT No. IN113920. We dedicate this work to our colleague and friend Bertrand Berche in the celebration of his 60th anniversary.

\section*{Appendix}
\setcounter{equation}{0}
\renewcommand{\theequation}{A.\arabic{equation}}
Here, we present the details of calculations of some quantities of interest presented in the main text. 
\subsection*{Calculation of Landau level spectrum and the expansion coefficients for the static regime}
First we determine the static Landau level spectrum. We start by writing the KeK-Y Hamiltonian equation (\ref{h0}) in the basis given in equation (\ref{ansatz}) which amounts to writing the Schr\"odinger equation $\mathcal{H}_0|\Psi_n^{\sigma\tau}\rangle=\sigma E^\tau_n|\Psi_n^{\sigma\tau}\rangle$ explicitly as
\begin{equation}
\hbar\omega_c\left(
\begin{array}{cccc}
0&\sqrt{n}&\Delta_0\sqrt{n} &0  \\
\sqrt{n}&0&0&\Delta_0\sqrt{n+1}  \\
\Delta_0\sqrt{n}&0&0&\sqrt{n+1}  \\
0&\Delta_0\sqrt{n+1}&\sqrt{n+1} &0  \\
\end{array}    
\right)\left(\begin{array}{c}
A_n^\tau \\
B_n^\tau \\
C_n^\tau \\ 
D_n^\tau 
\end{array}
\right)=E_n^\tau \left(\begin{array}{c}
A_n^\tau \\
B_n^\tau \\
C_n^\tau \\ 
D_n^\tau 
\end{array}
\right).
\end{equation}
Defining $\epsilon_n^\tau=E^\tau_n/\hbar\omega_c$ we get the secular equation:
\begin{equation}
\left|
\begin{array}{cccc}
-\epsilon_n^\tau&\sqrt{n}&\Delta_0\sqrt{n} &0  \\
\sqrt{n}&-\epsilon_n^\tau&0&\Delta_0\sqrt{n+1}  \\
\Delta_0\sqrt{n}&0&-\epsilon_n^\tau&\sqrt{n+1}  \\
0&\Delta_0\sqrt{n+1}&\sqrt{n+1} &-\epsilon_n^\tau  \\
\end{array} 
\right|=0,
\end{equation}
which in turn leads to the equation
\begin{equation}
(\epsilon_n^\tau)^4-(\epsilon_n^\tau)^2[(2n+1)(1+\Delta_0^2)]+n(n+1)(1-\Delta_0^2)^2=0.
\end{equation}
Completing the square, we get
\begin{equation}
[(\epsilon_n^\tau)^2-(n+1/2)(1+\Delta_0^2)]^2+n(n+1)(1-\Delta_0^2)^2=(n+1/2)^2(1+\Delta_0^2)^2,
\label{apA4}
\end{equation}
which we rewrite as
\begin{equation}
[(\epsilon_n^\tau)^2-(n+1/2)(1+\Delta_0^2)]^2+n(n+1)(1-2\Delta_0^2+\Delta_0^4)=n(n+1)(1+2\Delta_0^2+\Delta_0^4)+\frac{1}{4}(1+\Delta_0^2),
\label{apA5}
\end{equation}
that leads to the simpler expression
\begin{equation}
[(\epsilon_n^\tau)^2-(n+1/2)(1+\Delta_0^2)]^2=\frac{1}{4}[16n(n+1)\Delta_0^2+(1+\Delta_0^2)].
\label{apA6}
\end{equation}
We get then the result
\begin{equation}
\epsilon_n^\tau=\sqrt{\left(n+\frac{1}{2}\right)(1+\Delta_0^2)+\frac{1}{2}\tau\sqrt{16n(n+1)\Delta_0^2+(1+\Delta_0^2)^2}},\quad n = 1, 2, 3,\ldots\,.  
\end{equation}
To obtain the expansion coefficients for the eigenstate, we write down the corresponding set of equations
\begin{eqnarray}
\sqrt{n}(B_n^\tau+\Delta_0C_n^\tau)=\epsilon_n^{\tau}A_n^\tau,\nonumber\\
\sqrt{n}A_n^\tau+\Delta_0\sqrt{n+1}D_n^\tau=\epsilon_n^{\tau}B_n^\tau,\nonumber\\
\Delta_0\sqrt{n}A_n^\tau+\sqrt{n+1}D_n^\tau=\epsilon_n^{\tau}C_n^\tau,\nonumber\\
\sqrt{n+1}(\Delta_0B_n^\tau+C_n^\tau)=\epsilon_n^{\tau}D_n^\tau.
\label{apA8}
\end{eqnarray}
This set of equations is not linearly independent. We need to select three of them in order to find solutions for three parameters expressed in terms of the fourth. In addition, we need to treat separately the solutions for $\tau=\pm1$ as they have different limiting behavior when $\Delta_0\rightarrow0$. We show the explicit derivation for the $\tau=-1$ subspace, for which we need to use the set of coupled equations
\begin{eqnarray}\label{tauminus}
\sqrt{n}(B_n^-+\Delta_0C_n^-)=\epsilon_n^{-}A_n^-,\\
\Delta_0\sqrt{n}A_n^-+\sqrt{n+1}D_n^-=\epsilon_n^-C_n^-,\\
\sqrt{n+1}(\Delta_0B_n^{-}+C_n^-)=\epsilon_n^{-}D_n^{-},
\end{eqnarray}
which in the limit of vanishing Kekul\'e  parameter $\Delta_0\rightarrow0$ reduces to $\epsilon_n^-=\sqrt{n}$, $A^{-}_n,B^{-}_n\rightarrow1$ and $C^{-}_n,D^{-}_n\rightarrow0$. To solve the system, we apply the standard algebraic techniques as we show explicitly for $B^-_n$. Multiplication of equation (\ref{apA5}) by $\epsilon_n^-$ and substitution of equation (\ref{apA6}), give us
\begin{eqnarray}
&&\sqrt{n}(B_n^-+\Delta_0C_n^-)=\epsilon_n^{-}A_n^-,\\
&&\epsilon_n^-\Delta_0\sqrt{n}A_n^-+(n+1)(\Delta_0B_n^{-}+C_n^-)=(\epsilon_n^-)^2C_n^-,
\end{eqnarray}
which upon substitution of $C_n^-$ on (\ref{apA4}) can be further reduced to
\begin{equation}
B_n^-+\Delta_0\left[\frac{\epsilon_n^-\Delta_0\sqrt{n}A_n^-+(n+1)\Delta_0B_n^{-}}{(\epsilon^{-}_n)^2-(n+1)}\right]=\frac{\epsilon_n^{-}}{\sqrt{n}}A_n^-  ,  
\end{equation}
which is equivalent to
\begin{equation}
B_n^{-}\left[1+\frac{(n+1)\Delta^2_0}{(\epsilon^{-}_n)^2-(n+1)}\right]=\frac{\epsilon_n^{-}}{\sqrt{n}}\left[1-\frac{n\Delta^2_0}{(\epsilon^{-}_n)^2-(n+1)}\right]A_n^-,
\end{equation}
that leads to the result
\begin{equation}
B_n^{-}=\frac{\epsilon_n^{-}}{\sqrt{n}}\left[\frac{(\epsilon^{-}_n)^2-(n+1)-n\Delta_0^2}{(\epsilon^{-}_n)^2-(n+1)(1-\Delta_0^2)}\right]A_n^-\equiv b^-_n A^-_n.
\end{equation}
To obtain $C^-_n$, we use this result. First, we rewrite  equation~(\ref{apA8}) 
\begin{equation}
C_n^{-}=\frac{\Delta_0(\sqrt{n}\epsilon_n^{-}A^-_n+(n+1)B^-_n}{(\epsilon^{-}_n)^2-(n+1)},
\end{equation}
factoring out $A^-_n$, we  get
\begin{equation}
C_n^{-}=\frac{\Delta_0(\sqrt{n}\epsilon_n^{-}+(n+1)b^-_n}{(\epsilon^{-}_n)^2-(n+1)}A_n^-\equiv c^-_n A^-_n.
\end{equation}
To find $D^-_n$, we rewrite equation (\ref{apA5}) as
\begin{equation}
D_n^{-}=\frac{\epsilon_n^{-}C^-_n-\Delta_0\sqrt{n}A^-_n}{\sqrt{n+1}}.
\end{equation}
Using the result for $C^-_n$, we write this as
\begin{equation}
D_n^{-}=\frac{\Delta_0}{\sqrt{n+1}}\left(\frac{(\epsilon_n^{-})^2\sqrt{n}+\epsilon_n^-(n+1)b^-_n}{(\epsilon^-_n)^2-(n+1)}-\sqrt{n}\right)A^-_n,
\end{equation}
which is equivalent to
\begin{equation}
D_n^{-}=\Delta_0\sqrt{n+1}\left(\frac{\epsilon_n^-b^-_n+\sqrt{n}}{(\epsilon^-_n)^2-(n+1)}\right)A^-_n\equiv d^-_n A^-_n.
\end{equation}
The other coefficients are calculated in a similar fashion.
\subsection*{Calculation of the expansion coefficients via perturbation theory}
We now determine the expansion coefficients for the Floquet states as obtained from time- independent perturbation theory. The following results hold for $n\leqslant2$  since the $n=1$ Landau level couples with the $n=0$ level, which as we have found in the main text, is independent of the magnetic field and its coefficients do not follow the general result given in equation (\ref{ansatz}). We also give the leading order perturbation result within the off resonance scenario 
\begin{equation}\label{corrected}
|\Phi^{\tau\sigma}_n\rangle=|\Psi^{\tau\sigma}_n\rangle+\lambda\sum_{n'\neq n}\frac{\langle\Psi^{-\tau,-\sigma}_{n'}|(b+b^\dagger)|\Psi^{\tau\sigma}_n\rangle}{\sigma (E^\tau_n+E^{-\tau})-(n-n')\hbar\omega}|\Psi^{-\tau,-\sigma}_{n'}\rangle.    
\end{equation}
Using the static eigenstates (\ref{ansatz}), the matrix element  
\begin{equation}
\langle\Psi^{-\tau,-\sigma}_{n'}|b|\Psi^{\tau\sigma}_n\rangle ,   
\end{equation}
is found to be given by
\begin{equation}
\langle\Psi^{-\tau,-\sigma}_{n'}|b|\Psi^{\tau\sigma}_n\rangle=\frac{1}{N^\tau N^{-\tau}}(\sqrt{n-1}A^\tau_nA^{-\tau}_{n'}-\sqrt{n}(B^\tau_nB^{-\tau}_{n'}-C^\tau_nC^{-\tau}_{n'}-\sqrt{n+1}D^\tau_nD^{-\tau}_{n'})\delta_{n',n-1},    
\end{equation}
with $N^\tau=\sqrt{|A^\tau_n|^2+|B^\tau_n|^2+|C^\tau_n|^2+|D^\tau_n|^2}$.

\noindent We also obtain
\begin{equation}
\langle\Psi^{-\tau,-\sigma}_{n'}|b^\dagger|\Psi^{\tau\sigma}_n\rangle=\frac{1}{N^\tau N^{-\tau}}(\sqrt{n}A^\tau_nA^{-\tau}_{n'}-\sqrt{n+1}(B^\tau_nB^{-\tau}_{n'}-C^\tau_nC^{-\tau}_{n'}-\sqrt{n+2}D^\tau_nD^{-\tau}_{n'})\delta_{n',n+1},    
\end{equation}
upon substitution of these matrix elements in equation (\ref{corrected}), and performing the summations by means of the Kronecker symbols, we get
\begin{equation}\label{corrected}
|\Phi^{\tau\sigma}_n\rangle=|\Psi^{\tau\sigma}_n\rangle+\sum_sf^{\sigma\tau}_{n+s}|\Psi^{-\tau,-\sigma}_{n+s}\rangle.    
\end{equation}
where we can infer from the inspection that the coefficients are given by:
\begin{align}
&f^{\sigma\tau}_{n+s}= \left(\frac{e\mathcal{E}\ell_B}{\sqrt{2}}\right)\nonumber\\
&\times\frac{\sqrt{n-0.5(1-s)}A^\tau_nA^{-\tau}_{n+s}-\sqrt{n+0.5(1+s)}(B^\tau_nB^{-\tau}_{n+s}-C^\tau_nC^{-\tau}_{n+s})-\sqrt{n+1+0.5(1+s)}D^\tau_nD^{-\tau}_{n+s}}{\sigma(E^{-\tau}_{n+s}+E^\tau_n)-s\hbar\omega}.
 \end{align}

\ukrainianpart

\title[Взаємозв'язок між викривленнями Кекуле]%
{Взаємозв'язок між викривленнями Кекуле та лазерними полями у графені}
\author
{{А. Лопес\refaddr{label1},	{Ф. Мірелес\refaddr{label2}}}}
\addresses{
	\addr{label1}Вища політехнічна школа Побережжя, факультет фізики, природничих наук та математики, Кампус Густаво Галіндо
	30.5 км Віа Періметрал, 09-01-5863, Гуаякіль, Еквадор
	\addr{label2} Фізичний факультет, Центр нанотехнологій, Автономний національний університет Мехіко, 
	22800 Енсенада, Баха Каліфорнія, Мексика}

\makeukrtitle

\begin{abstract}
	Ми вивчаємо взаємозв'язок між структурним викривленням Кекуле у моношаровому графені та монохроматичним лазерним опромінюванням, фокусуючи увагу на довгохвильовому наближенні для структури рівнів Ландау. 
	Використовуючи дані щодо симетрії системи, в статичному режимі отримано точний аналітичний розв'язок для енергетичного спектру та його власних станів, що, у свою чергу, дозволяє нам знайти наближені вирази для поляризації зі ступенями свободи т. зв. ``долини'' та псевдоспіну.
	З'ясовано, що завдяки взаємодії ``долина-імпульс'' поляризація плоскої зони має чітко виражений магнітний відгук для двох компонентів $K$-зони. Завдяки цьому введення поля випромінювання змішує рівні Ландау. Показано, що налаштування системи на резонанс призводить до сильнішого відгуку поляризації плоскої зони у порівнянні з нерезонансним сценарієм.
	\keywords графен, рівні Ландау, викривлення Кекуле, теорія Флоке, поляризація
\end{abstract}


\lastpage
\end{document}